\documentclass[fina2,3l,leqno]{siamltex}
\usepackage{graphicx}
\usepackage{amsmath}
\usepackage{amsfonts}
\usepackage{amssymb}
\usepackage{enumerate}
\usepackage{subfigure}
\usepackage{color}
\usepackage{verbatim}
\usepackage{url}
\usepackage{alltt}   % for mostly 'verbatim'     
\usepackage{algorithm2e}
\usepackage{qtree}
\usepackage{mathrsfs}
\usepackage{eufrak}
\usepackage{yfonts}
\usepackage{qtree}

\usepackage{simplemargins}

\setleftmargin{1in}
\setrightmargin{1in}
\settopmargin{1in}
\setbottommargin{1in}

\newcounter{saveeqn}%

\DeclareMathAlphabet{\mathpzc}{OT1}{pzc}{m}{it}

\newtheorem{theo}{Theorem}[section]

\newtheorem{Defn}{Definition}[section]
\newtheorem{Prop}{Conjecture}[section]
\author{S.J. Gismondi\footnotemark[1], E.R. Swart\footnotemark[2]}
\title{Using  Matching to Detect Infeasibility of Some Integer Programs} 
\begin{document}

\renewcommand{\thefootnote}{\fnsymbol{footnote}}

\footnotetext[1]{\textit{University of Guelph, Canada, Email: \textup{\nocorr \texttt{gismondi@uoguelph.ca}}} (Corresponding Author)}
\footnotetext[2]{\textit{Kelowna, British Columbia, Canada, Email: \textup{\nocorr \texttt{ted.swart@shaw.ca}}}}
\footnotetext{Ted and I dedicate this paper to the late Pal Fischer (16/11/2016) - friend, colleague and mentor.}
%%%%%%%%%%%%%%%%%%%%%%%%%%%%%%%%%%%%%%%%%%%%%%%%
\maketitle

%We present a novel matching based heuristic algorithm intended to detect specially formulated infeasible IPs. It either detects an infeasible IP or exits undecided. It does not solve an IP.  We call it \textit{the triple overlay matching based closure algorithm} (the algorithm). Input to the algorithm is a $\{0,1\}$ IP whose constraints are a set of nested doubly stochastic subsystems together with a set of instance defining variables set at zero level. Its solution set is a subset of the set of $n!$ $n$x$n$ permutation matrices $P$, written as $n!$ $n^2$x$n^2$ block permutation matrices $Q$ each with block structure $P$. Output from the algorithm is a certificate of infeasibility, or an undecided IP and a set of variables deduced to be at zero level. Infeasible IPs may fail to be detected infeasible, while feasible IPs must fall in the undecided category. We present an application for the algorithm, a specially constructed $\{0,1\}$ IP model of the Hamilton tour decision problem, including empirical results. We successfully apply the algorithm to over 2,100 non-Hamiltonian graphs. No IPs fail that are not reported. We show how to model both the graph and subgraph isomorphism decision problems for input to the algorithm. We propose the algorithm can be developed for use with current solvers i.e. its output might help direct where and how to search for possible solutions.
\begin{abstract} A novel matching based heuristic algorithm designed to detect specially formulated infeasible $\{0,1\}$ IPs is presented. The algorithm's input is a set of nested doubly stochastic subsystems and a set $E$ of instance defining variables set at zero level.  The algorithm deduces additional variables at zero level until either a constraint is violated (the IP is infeasible), or no more variables can be deduced zero (the IP is undecided). All feasible IPs, and all infeasible IPs not detected infeasible are undecided. We successfully apply the algorithm to a small set of specially formulated infeasible $\{0,1\}$ IP instances of the Hamilton cycle decision problem. We show how to model both the graph and subgraph isomorphism decision problems for input to the algorithm.  Increased levels of nested doubly stochastic subsystems can be implemented dynamically. The algorithm is designed for parallel processing, and for inclusion of techniques in addition to matching.\end{abstract}

%\begin{keywords}
%cubic graph, 3-regular graph, snark, hamilton tour, hamilton cycle, decision problem, NP, coNP
%\end{keywords}

%\begin{AMS}
%05C45, 68R10, 90C05, 90C10
%\end{AMS}


\normalsize

\thispagestyle{empty}

%\pagenumbering{gobble}

\markboth{Swart \& Gismondi}{Detecting Infeasibile IPs via Matching} 

%%%%%%%%%%%%%%%%%%%%%%%%%%%%%%%%%%%5%%%%%%%%%%%%%
\section{Introduction}\label{section1} We present a novel matching based heuristic algorithm deigned to detect specially formulated infeasible $\{0,1\}$ IPs. It either detects an infeasible IP or exits undecided. It does not solve an IP. We call it \textit{the triple overlay matching based closure algorithm} (the algorithm). Input to the algorithm is an IP whose constraints are a set of nested doubly stochastic boolean subsystems \cite{sgsb16} together with a set $E$ of instance defining variables set at zero level. The IP's solution set is a subset of the set of $n!$ $n$x$n$ permutation matrices $P$, written as $n^2$x$n^2$ block permutation matrices $Q$ each with block structure $P$. The algorithm is a polynomial time search that deduces additional variables at zero level via matching until either a constraint is violated in which case the IP is infeasible, or we can go no further in which case the IP is undecided. If the IP is decided infeasible, a set of variables deduced to be at zero level can be used to test and display a set of violated constraints. If the IP is undecided, additional variables deduced zero can be added to $E$, and nothing more can be concluded. While some infeasible IPs may fail to be detected infeasible (not yet found), feasible IPs can only fall in the undecided category.\ 

In section \ref{section2} we present the generic IP required as input to the algorithm, and we view the set of all solutions of the IP as an $n^2$x$n^2$ block permutation matrix $Q$ whose components are $\{0,1\}$ variables. Each $n$x$n$ block $(u,i)$ is $n$x$n$ permutation matrix $P$ where block $(u,i)$ contains $p_{u,i}$ in position $(u,i)$. An instance is modelled by setting certain variables of $Q$ to zero level. In sections \ref{section3}, \ref{section4} and \ref{section5}, we present the algorithm, an application / matching model of the Hamilton cycle decision problem (HCP), empirical results and two conjectures. In section \ref{section6}, we present generalizations of the algorithm, matching models for both the graph and subgraph isomorphism decision problems, and other uses. We also propose more development. Its success, effectiveness and practicality can then be evaluated in comparison to other algorithms. We invite researchers to collaborate with us. Contact the corresponding author for FORTRAN code.\

The ideas presented in this paper originated from a polyhedral model of cycles not in graphs \cite{gs04}. At that time we thought about how to recognize the Birkhoff polytope as an image of a solution set of a compact formulation for non-Hamiltonian graphs. We've accomplished part of that goal in this paper. That is, the convex hull of all excluded permutations for infeasible IPs \textit{is} the Birkhoff polytope, and its easy to build a compact formulation from $E$. In this paper, over 2,100 non-Hamiltonian graphs ranging from 10 - 104 vertices are correctly decided as infeasible IPs. None failed that are not reported. Although counterexamples surely exist, we believe there is an insightful theory to be discovered that explains these early successes.\

\section{About Specially Constructed $\{0,1\}$ IPs and Terminology}\label{section2} Imagine a $\{0,1\}$ integer program modelled such that $P$ is a solution if and only if the integer program is feasible e.g. matching. Also imagine an arbitrary set of instance defining constraints of the form $p_{u,i} + p_{v,j} \le 1$. It's not obvious how to apply matching to help in its solution. Now imagine that we create a compact formulation whose solution set is isomorphic (i.e. equal under an orthogonal projection), where we convert each linear constraint into all of its instantiated discrete states via creation of a set of discrete $\{0,1\}$ variables. Then it becomes easy to exploit matching. Hence the algorithm.\

Code the IP above so that each of the instance defining constraints is a set of two distinct components of $P$ $\{p_{u,i} ,p_{v,j}\}$, interchangeably playing the role of a $\{0,1\}$ variable for which $\{p_{u,i},p_{v,j}\}=0$ if and only if $p_{u,i} = p_{v,j}=0$, or $p_{u,i}=1$ and $p_{v,j}= 0$, or $p_{u,i}=0$ and $p_{v,j}=1$. Create an instance of the IP by creating an instance of \textit{exclusion set $E$} whose elements are the set of these $\{p_{u,i},p_{v,j}\}$. If there exists $P$ satisfying $\{p_{u,i},p_{v,j}\}=0$ for all $\{p_{u,i},p_{v,j}\} \in E$, then $P$ is a solution of the IP. Otherwise $P$ satisfies $\{p_{u,i},p_{v,j}\}=1$ for at least one $\{p_{u,i},p_{v,j}\} \in E$ and $P$ is excluded from the solution set of the IP. We view elements of $E$ as coding precisely the set of permutation matrices excluded from the solution set of the IP. That is, $E$ excludes the union of sets of $(n-2)!$ $P$, each set satisfying $\{p_{u,i}, p_{v,j}\}$=1, for each $\{p_{u,i}, p_{v,j}\} \in E$. An example of the modelling technique needed to create $E$ is presented in section \ref{section4}, originally presented in \cite{gs04}. We exclude these permutation matrices by setting $\{p_{u,i},p_{v,j}\}=0$ for each $\{p_{u,i},p_{v,j}\} \in E$.\

The complement of exclusion set $E$, with respect to all $\{p_{u,i} ,p_{v,j}\}$ is called available set $V$. The IP is feasible if and only if there exists $P$ whose set of $\frac{n(n-1)}{2}$ distinct pairs of components $\{p_{u,i} ,p_{v,j}\}$ that satisfy $p_{u,i}=p_{v,j}=1$ and define $P$ are in $V$. $P$ is said to be covered by $V$ if there exists a subset of $\frac{n(n-1)}{2}$ $\{p_{u,i},p_{v,j}\} \in V$ such that $p_{u,i}=p_{v,j}=1$ defines $P$ and each $\{p_{u,i},p_{v,j}\}$ participates in $P$'s cover.\

\begin{Defn}
$Clos(E)$ (closed exclusion set $E$) is the set of all $\{p_{u,i},p_{v,j}\}$ not participating in any cover of any $P$.
\end{Defn}\

Note that $\{p_{u,i},p_{v,j}\} \in E$ is code for the set of $(n-2)!$ permutation matrices for which $p_{u,i}=p_{v,j}=1$. Clearly if $clos(E)$ is such that all $n!$ permutation matrices are accounted, then there is no $P$ covered by $V$ i.e. $V$ is empty.\  

\begin{Defn}
$Open(V)$ (open available set $V$) is the complement of $clos(E)$ w.r.t. all $\{p_{u,i},p_{v,j}\}$, the set of all $\{p_{u,i},p_{v,j}\}$ participating in a cover of at least one $P$.
\end{Defn}

\begin{theo}
The IP is infeasible if and only if $open(V)=\emptyset$.
\end{theo}

\textbf{System 1:}\
     $\,\,\sum_{i}p_{u,i}=1, u=1,2,...,n$\

     $\,\,\,\,\,\,\,\,\,\,\,\,\,\,\,\,\,\,\,\,\,\,\,\,\,\,\,\,\,\,\,\,\,\sum_{u}p_{u,i}=1, i=1,2,...,n$\
      
     $\,\,\,\,\,\,\,\,\,\,\,\,\,\,\,\,\,\,\,\,\,\,\,\,\,\,\,\,\,\,\,\,\,For\,\,all\,\,u,i=1,2,...,n$
      
     $\,\,\,\,\,\,\,\,\,\,\,\,\,\,\,\,\,\,\,\,\,\,\,\,\,\,\,\,\,\,\,\,\,\,\,\,\,\,\,\,\sum_{j\neq i}\{p_{u,i},p_{v,j}\} = p_{u,i}, v=1,2,...,n$, $v\neq u$.\

     $\,\,\,\,\,\,\,\,\,\,\,\,\,\,\,\,\,\,\,\,\,\,\,\,\,\,\,\,\,\,\,\,\,\,\,\,\,\,\,\,\sum_{v\neq u}\{p_{u,i},p_{v,j}\} = p_{u,i}, j=1,2,...,n$, $j\neq i$.\      

     $\,\,\,\,\,\,\,\,\,\,\,\,\,\,\,\,\,\,\,\,\,\,\,\,\,\,\,\,\,\,\,\,\,\,\,\,\,\,\,\,\{p_{u,i},p_{v,j}\} \in E\,\Rightarrow \,Assign\,\{p_{u,i},p_{v,j}\}=0.$\\      

      $\,\,\,\,\,\,\,\,\,\,\,\,\,\,\,\,\,\,\,\,\,\,\,\,\,\,\,\,\,\,\,\,\,p_{u,i}, \{p_{u,i},p_{v,j}\} \in \{0,1\}$\\

Visualize system 1 in the form of $n^2$x$n^2$ permutation matrix $Q$, $n^2$ blocks of $P$. Block $(u,i)$ contains $p_{u,i}$ in position $(u,i)$, the remaining entries in row $u$ and column $i$ being zero. The rest of the entries in block $(u,i)$ have the form $\{p_{u,i}, p_{v,j}\}$, $v\ne u, j\ne i$. It's assumed variables in $Q$ have been initialized by $E$. Henceforth, we present the algorithm in terms of matrix $Q$. See Figure \ref{Qmatter}, an example of the general form of matrix $Q$ for $n=4$. For $E=\emptyset$, the set of $n!$ $n$x$n$ permutation matrices each written as a $Q$ matrix i.e. in $n^2$x$n^2$ block form, is the set of integer extrema of the solution set of system 1\footnotemark[4]. See Figure  \ref{samplematter}, an example of an integer solution to system 1 in matrix $Q$ form, for $n=4$.\footnotetext[4]{An integer solution of system 1 exists if and only if it's an $n$x$n$ permutation matrix in $n^2$x$n^2$ block form \cite{gismo08}.}\

\begin{figure}[h!]
\begin{center}
\tiny
\setlength\arraycolsep{1pt}
$\left(\begin{array}{cccc|cccc|cccc|cccc} 

p_{11} & 0 & 0 & 0 & 0 & p_{12} & 0 & 0 & 0 & 0 & p_{13} & 0 & 0 & 0 & 0 & p_{14} \\
0&\{p_{11}p_{22}\} & \{p_{11}p_{23}\} & \{p_{11}p_{24}\} & \{p_{12}p_{21}\} & 0 & \{p_{12}p_{23}\} & \{p_{12}p_{24}\} & \{p_{13}p_{21}\} & \{p_{13}p_{22}\} & 0 & \{p_{13}p_{24}\} & \{p_{14}p_{21}\} & \{p_{14}p_{22}\} & \{p_{14}p_{23}\} & 0 \\ 
0 & \{p_{11}p_{32}\} & \{p_{11}p_{33}\} & \{p_{11}p_{34}\} & \{p_{12}p_{31}\} & 0 & \{p_{12}p_{33}\} & \{p_{12}p_{34}\} & \{p_{13}p_{31}\} & \{p_{13}p_{32}\} & 0 & \{p_{13}p_{34}\} & \{p_{14}p_{31}\} & \{p_{14}p_{32}\} & \{p_{14}p_{33}\} & 0 \\ 
0 & \{p_{11}p_{42}\} & \{p_{11}p_{43}\} & \{p_{11}p_{44}\} & \{p_{12}p_{41}\} & 0 & \{p_{12}p_{43}\} & \{p_{12}p_{44}\} & \{p_{13}p_{41}\} & \{p_{13}p_{42}\} & 0 & \{p_{13}p_{44}\} & \{p_{14}p_{41}\} & \{p_{14}p_{42}\} & \{p_{14}p_{43}\} & 0 \\ 
\\\hline
0 & \{p_{21}p_{12}\} & \{p_{21}p_{13}\} & \{p_{21}p_{14}\} & \{p_{22}p_{11}\} & 0 & \{p_{22}p_{13}\} & \{p_{22}p_{14}\} & \{p_{23}p_{11}\} & \{p_{23}p_{12}\} & 0 & \{p_{23}p_{14}\} & \{p_{24}p_{11}\} & \{p_{24}p_{12}\} & \{p_{24}p_{13}\} & 0 \\ 
p_{21} & 0 & 0 & 0 & 0 & p_{22} & 0 & 0 & 0 & 0 & p_{23} & 0 & 0 & 0 & 0 & p_{24} \\
0 & \{p_{21}p_{32}\} & \{p_{21}p_{33}\} & \{p_{21}p_{34}\} & \{p_{22}p_{31}\} & 0 & \{p_{22}p_{33}\} & \{p_{22}p_{34}\} & \{p_{23}p_{31}\} & \{p_{23}p_{32}\} & 0 & \{p_{23}p_{34}\} & \{p_{24}p_{31}\} & \{p_{24}p_{32}\} & \{p_{24}p_{33}\} & 0 \\ 
0 & \{p_{21}p_{42}\} & \{p_{21}p_{43}\} & \{p_{21}p_{44}\} & \{p_{22}p_{41}\} & 0 & \{p_{42}p_{33}\} & \{p_{22}p_{44}\} & \{p_{23}p_{41}\} & \{p_{23}p_{42}\} & 0 & \{p_{23}p_{44}\} & \{p_{24}p_{41}\} & \{p_{24}p_{42}\} & \{p_{24}p_{43}\} & 0 \\ 
\\\hline
0 & \{p_{31}p_{12}\} & \{p_{31}p_{13}\} & \{p_{31}p_{14}\} & \{p_{32}p_{11}\} & 0 & \{p_{32}p_{13}\} & \{p_{32}p_{14}\} & \{p_{33}p_{11}\} & \{p_{33}p_{12}\} & 0 & \{p_{33}p_{14}\} & \{p_{34}p_{11}\} & \{p_{34}p_{12}\} & \{p_{34}p_{13}\} & 0 \\ 
0 & \{p_{31}p_{22}\} & \{p_{31}p_{23}\} & \{p_{31}p_{24}\} & \{p_{32}p_{21}\} & 0 & \{p_{32}p_{23}\} & \{p_{32}p_{24}\} & \{p_{33}p_{21}\} & \{p_{33}p_{22}\} & 0 & \{p_{33}p_{24}\} & \{p_{34}p_{21}\} & \{p_{34}p_{22}\} & \{p_{34}p_{23}\} & 0 \\ 
p_{31} & 0 & 0 & 0 & 0 & p_{32} & 0 & 0 & 0 & 0 & p_{33} & 0 & 0 & 0 & 0 & p_{34} \\
0 & \{p_{31}p_{42}\} & \{p_{31}p_{43}\} & \{p_{31}p_{44}\} & \{p_{32}p_{41}\} & 0 & \{p_{32}p_{43}\} & \{p_{32}p_{44}\} & \{p_{33}p_{41}\} & \{p_{33}p_{42}\} & 0 & \{p_{33}p_{44}\} & \{p_{34}p_{41}\} & \{p_{34}p_{42}\} & \{p_{34}p_{43}\} & 0 \\ 
\\\hline
0 & \{p_{41}p_{12}\} & \{p_{41}p_{13}\} & \{p_{41}p_{14}\} & \{p_{42}p_{11}\} & 0 & \{p_{42}p_{13}\} & \{p_{42}p_{14}\} & \{p_{43}p_{11}\} & \{p_{43}p_{12}\} & 0 & \{p_{43}p_{14}\} & \{p_{44}p_{11}\} & \{p_{44}p_{12}\} & \{p_{44}p_{13}\} & 0 \\ 
0 & \{p_{41}p_{22}\} & \{p_{41}p_{23}\} & \{p_{41}p_{24}\} & \{p_{42}p_{21}\} & 0 & \{p_{42}p_{23}\} & \{p_{42}p_{24}\} & \{p_{43}p_{21}\} & \{p_{43}p_{22}\} & 0 & \{p_{43}p_{24}\} & \{p_{44}p_{21}\} & \{p_{44}p_{22}\} & \{p_{44}p_{23}\} & 0 \\ 
0 & \{p_{41}p_{32}\} & \{p_{41}p_{33}\} & \{p_{41}p_{34}\} & \{p_{42}p_{31}\} & 0 & \{p_{42}p_{33}\} & \{p_{42}p_{34}\} & \{p_{43}p_{31}\} & \{p_{43}p_{32}\} & 0 & \{p_{43}p_{34}\} & \{p_{44}p_{31}\} & \{p_{44}p_{32}\} & \{p_{44}p_{33}\} & 0 \\ 
p_{41} & 0 & 0 & 0 & 0 & p_{42} & 0 & 0 & 0 & 0 & p_{43} & 0 & 0 & 0 & 0 & p_{44} \\
\end{array}\right)$
\normalsize
\caption{General Form of Matrix $Q$, $n=4$.}
\label{Qmatter}
\end{center}
\end{figure}

\begin{figure}[h!]
\begin{center}
\small
$\left(\begin{array}{cccc|cccc|cccc|cccc}
           0 & 0 & 0 & 0 & 0 & 0 & 0 & 0 & 0 & 0 & 1 & 0 & 0 & 0 & 0 & 0\\
           0 & 0 & 0 & 0 & 0 & 0 & 0 & 0 & 1 & 0 & 0 & 0 & 0 & 0 & 0 & 0\\
           0 & 0 & 0 & 0 & 0 & 0 & 0 & 0 & 0 & 1 & 0 & 0 & 0 & 0 & 0 & 0\\
           0 & 0 & 0 & 0 & 0 & 0 & 0 & 0 & 0 & 0 & 0 & 1 & 0 & 0 & 0 & 0\
           
\\\hline           
           0 & 0 & 1 & 0 & 0 & 0 & 0 & 0 & 0 & 0 & 0 & 0 & 0 & 0 & 0 & 0 \\
           1 & 0 & 0 & 0 & 0 & 0 & 0 & 0 & 0 & 0 & 0 & 0 & 0 & 0 & 0 & 0 \\
           0 & 1 & 0 & 0 & 0 & 0 & 0 & 0 & 0 & 0 & 0 & 0 & 0 & 0 & 0 & 0 \\
           0 & 0 & 0 & 1 & 0 & 0 & 0 & 0 & 0 & 0 & 0 & 0 & 0 & 0 & 0 & 0\
           
\\\hline
           0 & 0 & 0 & 0 & 0 & 0 & 1 & 0 & 0 & 0 & 0 & 0 & 0 & 0 & 0 & 0\\
           0 & 0 & 0 & 0 & 1 & 0 & 0 & 0 & 0 & 0 & 0 & 0 & 0 & 0 & 0 & 0\\
           0 & 0 & 0 & 0 & 0 & 1 & 0 & 0 & 0 & 0 & 0 & 0 & 0 & 0 & 0 & 0\\
           0 & 0 & 0 & 0 & 0 & 0 & 0 & 1 & 0 & 0 & 0 & 0 & 0 & 0 & 0 & 0\

\\\hline
           0 & 0 & 0 & 0 & 0 & 0 & 0 & 0 & 0 & 0 & 0 & 0 & 0 & 0 & 1 & 0\\
           0 & 0 & 0 & 0 & 0 & 0 & 0 & 0 & 0 & 0 & 0 & 0 & 1 & 0 & 0 & 0\\
           0 & 0 & 0 & 0 & 0 & 0 & 0 & 0 & 0 & 0 & 0 & 0 & 0 & 1 & 0 & 0\\
           0 & 0 & 0 & 0 & 0 & 0 & 0 & 0 & 0 & 0 & 0 & 0 & 0 & 0 & 0 & 1\

\end{array}\right)$
\normalsize
\caption{An Integer Solution to System 1 in Matrix $Q$ Form, $n=4$.}
\label{samplematter}
\end{center}
\end{figure}

\newpage

\section{About Triple Overlay Matching Based Closure}\label{section3} We first present an overview of the algorithm, followed by the formal algorithm. Let $E$ be given. Encode $Q$ and create $V$.\ 

\subsection{Overview of the Triple Overlay Matching Based Closure Algorithm}\label{section31} Rather that search for the existence of $P$ covered by $V$, we attempt to shrink $V$ so that $\{p_{u,i},p_{v,j}\} \in V$ if and only if $\{p_{u,i},p_{v,j}\}$ participates in a cover of at least one $P$. The algorithm deduces which $\{p_{u,i},p_{v,j}\} \in V$ do not participate in any cover of any $P$, removes them from $V$, and adds them to $E$. Its success depends upon whether or not it's true that for infeasible IPs, when we initialize $Q$ via $E$, it's sufficient to deduce $open(V)=\emptyset$\footnotemark[5]. While it's impossible for a feasible IP to yield $open(V)=\emptyset$, infeasible IPs cause the algorithm to either deduce infeasibility or exit undecided. We say undecided because although we deduce \textit{some} of these $\{p_{u,i},p_{v,j}\} \in V$ that do not participate in any cover of any $P$, it's not known if we deduce \textit{all} of these $\{p_{u,i},p_{v,j}\}$. Brief details about how the algorithm deduces variables at zero level in every solution of the IP now follow.\footnotetext[5]{In earlier work \cite{sgsb16}, we create an equivalence class, the set of all possible $V$'s none of which cover any $P$, whose class representative is $\emptyset$.}\

The algorithm systematically tests a set of necessary conditions assuming a feasible IP each time a $q_{u,i,v,j}$ is set at unit level. That is, if $p_{u,i}=p_{v,j}$=1, blocks $(u,i)$ and $(v,j)$ are assumed to cover a match, a necessary condition for the existence of a $n^2$x$n^2$ block permutation matrix solution of the IP. But rather than test for a match covered by these two blocks, we exhaust all choices of a third variable common to these blocks, set at unit level, and test for the existence of a match covered by all three blocks. After exhausting all possible choices of a third\footnotemark[9] variable, if no match exists, the given $q_{u,i,v,j}$ variable is deduced zero. Otherwise we conclude nothing. In both cases we continue on to the next variable not yet deduced zero. Eventually no more variables can be deduced zero, none of the constraints appear violated and the IP is undecided; or enough of the variables are deduced zero such that a constraint is violated and the IP is infeasible.\footnotetext[9]{Hence the term triple overlay. Every variable not deduced zero participates in a match in an overlay of three blocks of $Q$. There exists quadrupal, quintuple overlay through to exhaustion where the algorithm tests factorial numbers of $n$ ($n-2$ is sufficient) overlays for a match.}\

\subsection{The Triple Overlay Matching Based Closure Algorithm}\label{section32} Interchangeably associate matrix $Q$ with a $\{0,1\}$ matrix that has entries at zero level where matrix $Q$ has entries at zero level, and unit entries where matrix $Q$ has $p_{u,i}$ or $\{p_{u,i},p_{v,j}\}$ entries. We'll now reference $\{0,1\}$ variables $q_{u,i,u,i}$ and $q_{u,i,v,j}$. A unit entry in the $u^{th}$ row and $i^{th}$ column of non-empty block $(u,i)$ represents variable $p_{u,i}$. The remaining unit entries in the $v^{th}$ row and $j^{th}$ column of block $(u,i)$ with $u\ne v$ and $i \ne j$ can be regarded as representing $p_{v,j}$ variables (which is what they really do represent in the case of a $\{0,1\}$ solution) and, they can also be regarded as representing $\{p_{u,i},p_{v,j}\}$ variables. We think of this associated matrix in terms of patterns in $Q$ that cover $n^2$x$n^2$ block permutation matrices, and then we'll exploit matching.\

\begin{Defn}
Match(*) is a logical function. Input * is an $n$x$n$ $\{0,1\}$ matrix. Row labels are viewed as vertices 1 through $n$ (set $A$), and column labels are viewed as vertices $n+1$ through $2n$ (set $B$). Match(*) returns TRUE if there exists a match between $A$ and $B$. Otherwise Match(*) returns FALSE.
\end{Defn}

\begin{Defn}
Overlay(*$_1$,*$_2$) is a binary AND function applied component-wise to two $n$x$n$ $\{0,1\}$ matrices. Its output is a $\{0,1\}$ matrix. We loosely use the terms double and triple overlay in place of Overlay(*$_1$,*$_2$) and Overlay(Overlay(*$_1$,*$_2$),*$_3$) etc.  
\end{Defn}

\begin{Defn}
$Check\_RowsColumns\_Q$ is a routine that returns TRUE if a row or column in matrix $Q$ is all 0, in which case the algorithm terminates and the graph is deduced infeasible. Otherwise $Check\_RowsColumns\_Q$ returns FALSE. In our FORTRAN implementation of the algorithm, before testing for termination, we also implement Boolean closure within and between blocks in $Q$. This efficiently deduces some of the non-zero components of $Q$ to be at zero level and we note significant speed increases.\end{Defn}

Note that Boolean closure in $Check\_RowsColumns\_Q$ can be replaced by LP. Temporarily set a non-zero component of matrix $Q$ to unit level and check for infeasibility subject to doubly stochastic constraints of matrix $Q$. Infeasibility implies that the component can be set to zero level.\

Whenever the algorithm exits undecided, then for every non-zero $q_{u,i,v,j}$, there exists a match in a triple overlay of blocks $(u,i)$, $(v,j)$ and at least one $(w,k)$ block. The IP is then not deduced infeasible and we call the corresponding matrix $Q$ the non-empty triple overlay closure of the IP. Otherwise the algorithm exits and the IP is deduced infeasible i.e. $open(V)$ is deduced to be empty.\ 

%\newpage

%%%%%%%%%%%%%%%%%%%%%%%%%% 
\begin{algorithm}
%\SetLine
\KwIn{\{$open(V)\leftarrow V$, Q$\}$}
\KwOut{$\{open(V)$, decision$\}$}
\textbf{if}\textit{{$\,Check\_RowsColumns\_Q$}} {\textbf{$\,$EXIT} $\{open(V)$, infeasible$\}$}\;

\textbf{CONTINUE TRIPLE CLOSURE}\;
  $oldQ \leftarrow Q$\;
   \For{$u,i=1,2,...,n$; and $q_{u,i,u,i}\ne 0$}
    {\If{$\tilde{}$Match$(Q(u,i))$}
         {$q_{u,i,u,i}\leftarrow 0$\;
            \For{$v,j=1,2,...,n$; and $u \ne v$ and $i\ne j$ and $q_{u,i,v,j} \ne 0$}
                  {$q_{u,i,v,j}\leftarrow 0$; $q_{v,j,u,i}\leftarrow 0$; $open(V)\leftarrow open(V)\backslash \{p_{u,i},p_{v,j}\} \backslash \{p_{v,j},p_{u,i}\}$\;
                  }
                   \textbf{if}\textit{{$\,Check\_RowsColumns\_Q$}} {\textbf{$\,$EXIT} $\{open(V)$, infeasible$\}$}\;
                   $\Rightarrow$ \textbf{NEXT \textit{i}}\;
          }
   \For{$v,j=1,2,...,n$; and $u\ne v$ and $i\ne j$ and $Q(u,i)_{v,j} \ne 0$}
    {\If{$\tilde{}$Match$($Overlay$(Q(u,i), Q(v,j)))$}
            {$q_{u,i,v,j}\leftarrow 0$; $q_{v,j,u,i}\leftarrow 0$;  $open(V)\leftarrow open(V)\backslash \{p_{u,i},p_{v,j}\} \backslash \{p_{v,j},p_{u,i}\}$\;
              $\Rightarrow$ \textbf{NEXT \textit{j}}\;}

        {\textit{DoubleOverlay} $\leftarrow$ \textit{Overlay}($Q(u,i), Q(v,j)$)\;
\textbf{TRIPLE CLOSURE}\;
          \For{$w,k=1,2,...,n$; and $u\ne w \ne v$, $i\ne k \ne j$ and DoubleOverlay$_{w,k}\ne 0$}
                 {\If{$\tilde{}$Match$($Overlay$($DoubleOverlay, $Q(w,k)))$} 
                      {\textit{DoubleOverlay}$_{w,k}\leftarrow 0$\;
                       $\Rightarrow$ \textbf{TRIPLE CLOSURE}\;
                      }     
                 }
    \If{$\tilde{}$Match$($DoubleOverlay$)$}
       {$q_{u,i,v,j}\leftarrow 0$; $q_{v,j,u,i}\leftarrow 0$;  $open(V)\leftarrow open(V)\backslash \{p_{u,i},p_{v,j}\} \backslash \{p_{v,j},p_{u,i}\}$\;
       }         
         } 
     }
   }
   
\textbf{if}\textit{$\,oldQ \ne Q$} {\textbf{$\Rightarrow$ CONTINUE TRIPLE CLOSURE}}\;
\textbf{EXIT} $\{open(V)$, undecided$\}$\;
\
\caption{The Triple Overlay Matching Based Closure Algorithm}
\label{alg:alg}
\end{algorithm}

\newpage

\section{Application to the HCP}\label{section4} Let $G$ be an $n+1$ vertex graph also referenced by its adjacency matrix. We model the HCP for simple, connected 3-regular graphs, as do others \cite{fhr15,hay15,hay2}, called the 3HCP.\

\subsection{Background Information and Classification of 3-Regular Graphs}\label{section41} The 3HCP is a well known decision problem and is \textbf{NP}-complete \cite{gjt76}. $G$ is 3-colourable (edge) if $G$ is Hamiltonian, and since 3-regular graphs are either 3 or 4-colourable, it follows that if $G$ is 4-colourable then $G$ is non-Hamiltonian. These graphs were initially studied by Peter Tait in the 1880s (named Snarks by Martin Gardner in 1976). Tait conjectured (1884) that every 3-connected, planar, 3-regular graph has a Hamilton cycle, later disproved by Tutte in 1946 via construction of a 46 vertex counterexample. This was a significant conjecture, and had it been true, it implied the famous 4-colour theorem. These ideas are summarized in the figure below.\

\begin{figure}[h!]
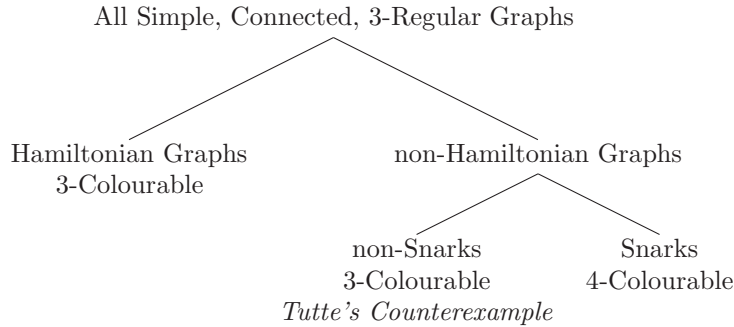

\begin{center}
\leaf{Hamiltonian Graphs\\ 3-Colourable}
\leaf{non-Snarks \\ 3-Colourable \\ \textit{Tutte's Counterexample}}
\leaf{Snarks \\ 4-Colourable}
\branch{2}{non-Hamiltonian Graphs\\
}
\branch{2}{All Simple, Connected, 3-Regular Graphs}
\qobitree
\end{center}
\caption{Classification of Simple, Connected, 3-Regular Graphs} \label{treeUEP}  
\end{figure} 

%\newpage

\subsection{The Matching Model of the HCP}\label{section42} Regard paths of length $n+1$ that start and stop at the same vertex and pass through every vertex, as directed graphs on $n+1$ vertices. For undirected graphs, every cycle is accompanied by a counter-directed companion cycle. No matter that $G$ is Hamiltonian or non-Hamiltonian, assign vertex $n+1$ as the origin and terminal vertex for all cycles, and assign each directed Hamilton cycle to be in correspondence with each $n$x$n$ permutation matrix $P$ where $p_{u,i}=1$ if and only if the $i^{th}$ arc in a cycle enters vertex $u$. We encode each cycle as a permutation of vertex labels. For example, the path sequence \{6,2,4,3,5,1,6\} is code for the first arc enters vertex 2, the second arc enters vertex 4 and so on. Since $p_{n+1,n+1}=1$ for all cycles by definition, it's sufficient to code cycles as $n$x$n$ permutation matrices. Note that an arc is directed, and an edge is undirected i.e. the pair of arcs $(u,i)$ \& $(i,u)$ is the edge $(u,i)$. Unless otherwise stated, all graphs are simple, connected and 3-regular.\

We next encode graph instance $G$ by examining $G$'s adjacency matrix, adding to $E$ all pairs of components of $P$, $\{p_{u,i} ,p_{v,j}\}$ that encode paths of length $j-i$ $(j>i)$ from vertex $u$ to vertex $v$ in cycles not in $G$. This encodes precisely the set of cycles not in $G$ i.e. every cycle not in $G$ uses at least one arc not in $G$.\

See the algorithm (How to Initialize Exclusion Set $E$) below and recall that $G$ is connected. For arc $(u,v)$ not in $G$ we can assign $\{p_{u,i},p_{v,i+1}\} \in E$. But we also compute additional $\{p_{u,i},p_{v,i+m}\}$ whenever it's possible to account for no paths of length $m$ in $G$, from vertex $u$ to vertex $v$. We do this by implementing Dijkstras algorithm with equally weighted arcs to find minimal length paths between all pairs of vertices, coded to return $m=n+1$ if no path exists. We account for all paths of length one not in $G$ (arcs not in $G$), and, all paths of length two not in $G$ by temporarily deleting the arc between adjacent vertices.\

Begin as follows. If $u$ is adjacent to $v$ then temporarily delete arc ($u$,$v$) and apply Dijkstras algorithm to discover a minimal path of length $m>1$, a simple `speed-up'. No paths of length $k$ can exist, $k=1,...,m-1$ and $\{p_{u,i},p_{v,i+k}\}$ are discovered that 1) for $k=1$ and $u$ not adjacent to $v$ correspond with arcs in cycles not in $G$, and 2) for $k>1$ correspond with paths of length $k$ in cycles not in $G$. Accounting for all arcs not in $G$ is sufficient to model precisely all cycles not in $G$, and we account for paths in cycles not in $G$ to bolster $E$.\

Two special cases arise. \textit{Case 1. Last arc in cycle}: Recall that every $n+1^{th}$ arc in a cycle enters vertex $n+1$ by definition. Therefore observe arcs ($u$,$n+1$) not in $G$, temporarily deleted or otherwise, noting how corresponding sets of cycles not in $G$ can be encoded by permutation matrices for which the $n^{th}$ arc in a cycle enters vertex $u$ i.e. $p_{u,n}=1$. This is the case for $k$=1, and $u$ not adjacent to $v$ when Dijkstras algorithm returns $m=2$. If Dijkstras algorithm returns $m=3$, then again for $k$=1 and if $u$ is not adjacent to $v$ set $p_{u,n}=1$, and for $k$=2, no paths of length two exist and these sets of cycles not in $G$ can be encoded by permutation matrices for which the $n-1^{th}$ arc in a cycle enters vertex $u$ i.e. $p_{u,n-1}=1$. Continuing in this way, encode all possible $n+1-k^{th}$ arcs in cycles not in $G$, in paths of length $k$ not in $G$, to enter vertex $u$ i.e. $p_{u,n+1-k}=1$, $k=1,2,...,m-1$. \textit{Case 2. First arc in cycle}: Recall every first arc in every cycle exits vertex $n+1$. Observe and code all arcs ($n+1$,$v$) in cycles not in $G$, in paths of length $k$ not in $G$ by coding all possible $k^{th}$ arcs to enter vertex $v$ i.e. $p_{v,k}$=1, $k=1,2,...,m-1$.\

For the general case, an exclusion set can be constructed by noting that a cycle not in $G$ uses at least one arc not in $G$ e.g. $(u,v)$. The complete set of permutation matrices corresponding to these cycles not in $G$ are characterized by $\{\{p_{u,l}, p_{v,l+1}\}$, $l=1,2,...,n-1\}$, added to $E$. By indexing $l$, arc $(u,v)$ can play the role of {($n$-1)} sequence positions in disjoint sets of cycles not in $G$. Considering all $\mathcal{O}(n^2)$ arcs not in $G$, each playing the role of all $\mathcal{O}(n)$ possible sequence positions, it's possible to construct the set of permutation matrices corresponding to the set of cycles not in $G$ accounted by the union of $\mathcal{O}(n^3)$ $\{p_{u,l},p_{v,l+1}\}$, added to $E$. We generalize this idea via Dijkstras algorithm and account for some sets of paths of length $k$ not in $G$.\

Recall that $G$ is strongly connected. But if an arc is temporarily deleted, it's possible for no path to exist between a given pair of vertices. This useful information indicates that an arc is essential under the assumption of the existence of a Hamilton cycle that uses this arc. In case 1, this implies that a particular $p_{u,n}$ is necessary, and by integrality must be at unit level in every assignment of variables, assuming the graph is Hamiltonian (until deduced otherwise, if ever). Thus all other $P$'s in the same row (and column) can be set at zero level. This is accounted for when we initialize $E$. Recall that $m=n+1$ in the case that Dijkstras algorithm returns no minimal path. The $k$ loop appends the necessary set of $\{p_{v,j},p_{u,n+1-k}\}$ to $E$ effectively setting variables in blocks ($u$,1,) through ($u$,$n-1$) at zero level. When implemented in the algorithm, $p_{u,n}$ must attain unit level via double stochastity, and this implies that the other $P$'s in the same column are deduced to be at zero level. Similarily for Case 2. In the general case, it's also possible for no path to exist between a given pair of vertices ($u$,$v$) (when an arc is temporarily deleted). Under the assumption of the existence of a Hamilton cycle, this arc is essential and can play the role of sequence position 2 through $n$-1 and so in each case, all complementary row (and column) $\{p_{u,i},p_{v,j}\}$ are assigned to $E$. When implemented, a single $\{p_{u,i},p_{v,j}\}$ variable remains in each row and therefore is  equated with that block's $p_{u,i}$ variable via `scaled' double stochastity \textit{within} the block i.e. rows and columns in the block sum to $p_{u,i}$. Complementary $\{p_{u,i},p_{v,j}\}$ variables in the corresponding column are therefore set 0 in each block. Thus essential arcs also contribute to new information by adding their complementary row / column $\{p_{u,i},p_{v,j}\}$ to $E$.\

Finally, encode $E$ into matrix $Q$ i.e. assign $q_{u,i,v,j}=0$ for each $\{p_{u,i},p_{v,j}\} \in$ $E$, and then create $V$.\

%%%%%%%%%%%%%%%%%%%%%%%%%% 
\begin{algorithm}
%\SetLine
\KwIn{\{Arc Adjacency matrix for $G$\}}
\KwOut{\{$E$\}}
$E\leftarrow\emptyset$\;
Case 1: \For{$u=1,2,...,n$}{
$Arc \leftarrow G(u,n+1); G(u,n+1) \leftarrow 0$; $m \leftarrow$ DijkstrasAlgorithm($G$,$u$,$n+1$)\;
\For{$k=Arc+1,Arc+2,...,m-1$}{
$E\leftarrow E\cup \{p_{v,j},p_{u,n+1-k}\}, v=1,2,...,n, v\neq u; j=1,2,...,n, j\neq n+1-k$\;
}
$G(u,n+1) \leftarrow Arc$\;
}
Case 2: \For{$v=1,2,...,n$}{
$Arc \leftarrow G(n+1,v); G(n+1,v) \leftarrow 0$; $m \leftarrow$ DijkstrasAlgorithm($G$,$n+1$,$v$)\;
\For{$k=Arc+1,Arc+2,...,m-1$}{
$E\leftarrow E\cup \{p_{v,k},p_{u,i}\}, u=1,2,...,n, u\neq v; i=1,2,...,n, i\neq k$\;
}
$G(n+1,v) \leftarrow Arc$\;
}
General Case. \For{$u=1,2,...,n$}{
\For{$v=1,2,...,n; v\neq u$}{
$Arc \leftarrow G(u,v); G(u,v) \leftarrow 0$; $m \leftarrow$ DijkstrasAlgorithm($G$,$u$,$v$)\;
\For{$k=Arc+1,Arc+2,...,m-1$}{
$E\leftarrow E\cup \{p_{u,l},p_{v,l+k}\}, l=1,2,...,n-k$\;
}
$G(u,v) \leftarrow Arc$\;
}
}
\textbf{EXIT} \{$E $\} \;
\
\caption{How to Initialize Exclusion Set $E$}
\label{alg:ExcSets}
\end{algorithm}

\newpage
\section{Empirical Results and Two Conjectures}\label{section5} Table \ref{tab1} below lists some details of 25 applications (all 3-regular graphs) of the algorithm. Table \ref{tab2} below lists some details of 20 applications (mostly 3-regular graphs) of an earlier version of the matching based closure algorithm called the WCA\footnotemark[7] \cite{sgsb16} (a subset from over 2,100 applications). For both algorithms, all of the graphs are decided non-Hamiltonian and no application of either algorithm to any other graphs failed that are not reported.\

\subsection{Empirical Results}\label{section51} In both tables, heading \# $p_{u,i} (|V|)$ is the count of non-zero $p_{u,i}$ variables and the size of initial available set $V$ (the number of non-zero $q_{u,i,v,j}$ components in $Q$) after initializing $E$, before implementing the algorithm. Note $p_{u,i}=q_{u,i,u,i}$. We only count $q_{u,i,v,j}$ $i<j$ (distinct $q_{u,i,v,j}$).\

In Table \ref{tab2}, heading $|open(V)| \le$ refers to an upper bound on  $|open(V)|$ for 11 selected graphs, each modified to include the cycle $1-2-...-n-(n+1)-1$, simply to observe $open(V)$. Two of these graphs are also hypohamiltonian. The count in parentheses is an upper bound on $|open(V)|$ after removing a vertex and re-running the WCA.\

\subsection{Two Conjectures}\label{section52} \

\begin{Prop}
\label{con1}
Polynomial sized proof of membership of all $\frac{n^2(n-1)^2}{2}$ distinct $\{p_{u,i},p_{v,j}\} \in E$ exists for all simple, connected, 3-regular, non-Hamiltonian graphs.
\end{Prop}\

\begin{Prop}
\label{con2}
Triple overlay matching based closure deduces $open(V)=\emptyset$ for all simple, connected, 3-regular, non-Hamiltonian graphs.
\end{Prop}\

\footnotetext[7]{The WCA is a breadth-first closure, exhausting the middle $v,j$ loop before returning to label CONTINUE TRIPLE CLOSURE. It's followed by triple closure, also applied breadth-first, exhausting the interior $w,k$ loop before returning to label TRIPLE CLOSURE. Many more applications of Boolean closure across all of $Q$ at many more intermediate steps are also implemented, unlike triple overlay matching based closure as we have presented (although these checks can also be included). Block overlays are also restricted to be of the form $Q(u,i)$ and $Q(v,j)$, $i<j$. In this way we can solve problems in the 50-100 vertex range. The WCA is designed to be parallelized, and the FORTRAN code is written for distributed computing.}

\newpage

%%%%%%%%%%%%%%%%%%%%%%%%%%%%%%%%%%%%%%%%%%%%%%%%%%
\begin{table}[h!]
\caption{Applications of the Triple Overlay Matching Based Closure Algorithm} %title of the table 
\label{tab1}
{\centering     % centering table 

\begin{tabular}{lcc}  % creating eight columns 
Name of Graph & \# Vertices (All 3-Regular Graphs) & \# $p_{u,i} (|V|)$\\ [1ex]    
\hline                % inserts single-line 
Petersen Snark & 10 & 57 (858)\\
3 Flower Snarks & 12, 20, 28 & 87 (2,199), 271 (26,380), 567 (126,128)\\  
Tietzs Snark & 12 & 87 (2,257)\\  
2 Blanusa Snarks & 18, 26 & 223 (16,630), 495 (88,968)\\  
House of Graphs\cite{bcgm2013} & 18 & 219 (16,262)\\
A Loupekine Snark & 22 & 345 (43,719)\\  
A Goldberg Snark & 24 & 419 (65,711)\\  
10 House of Graphs\cite{bcgm2013} & 26 & $\approx {500}$ ($\approx {1,000,000}$)\\  
Jan Goedgebeur Snark & 26 & 497 (89,188)\\  
2 Celmins-Swart Snarks & 26 & 505 (96,175), 509 (95,543)\\  
House of Graphs\cite{bcgm2013} & 28 & 583 (129,018)\\ 
A Double Star Snark & 30 & 691 (187,398)\\  
[1ex]    % [1ex] adds vertical space 
\hline   % inserts single-line 
\end{tabular} 

}\
\end{table}

%\newpage
%%%%%%%%%%%%%%%%%%%%%%%%%%%%%%%%%%%%%%%%%%%%%%%%%
\begin{table}[h!] 

\caption{Applications of A Matching Based Closure Algorithm (WCA) \cite{sgsb16}} %title of the table 
\label{tab2}
{\centering     % centering table 

\begin{tabular}{lccc}  % creating eight columns 
Name of Graph & \# Vertices (Edges) & \# $p_{u,i} (|V|)$ & $|open(V)|\le$  \\ [1ex]    
\hline                % inserts single-line 
Petersen Snark & 10 (15) & 57 (858) &  792\\  
Herschel Graph & 11 (18) & No 2-Factor &  1,980\\
A Kleetope & 14 (36) & 147 (8,166) &  5,809\\  
$^1$Matteo\cite{Leder16Other}\footnotemark[8] & 20 (30) & 275 (26,148) &  27,093\\
$^1$Coxeter & 28 (42) & 597 (136,599) & 135,453(1,241)$^2$\\  
House of Graphs \#3337 Snark\cite{bcgm2013} & 34 (51) & 897 (308,234) & 335,697\\
Zamfirescu Snark & 36 (54) & 983 (363,987) & 7,749\\
$^1$Barnette-Bosák-Lederberg\footnotemark[8] & 38 (57) & 1077 (440,318) & 96,834\\
$^1$A Hypohamiltonian & 45 (70) & 1,656 (1,109,738) & 296,668 (29,724)$^2$\\
$^1$Tutte\footnotemark[8] & 46 (69) & 1,649 (1,060,064) & 382,400\\
A Grinberg Graph & 46 (69) & 1,737 (1,204,722) & - Not run yet -\\  
$^1$Georges\footnotemark[8] & 50 (75) & 2037 (1,701,428?) & - Not run yet -\\
Szekeres Snark & 50 (75) & 2045 (1,718,336) & - Not run yet -\\
Watkins Snark & 50 (75) & 2051(1,708,987) & - Not run yet -\\
$^1$Ellingham-Horton & 54 (81) & 2,315 (2,135,948)  & 1,045,041\\
Thomassen & 60 (99) & 3,105 (4,071,600) & - Not run yet -\\  
Meredith & 70 (140) & 4,221 (7,526,996) & - Not run yet -\\
A Flower Snark & 76 (114) & 4,851 (9,720,420) & - Not run yet -\\  
$^1$Horton\footnotemark[8] & 96 (144) & 8,205 (29,057,118) & - Not run yet -\\  
A Goldberg Snark &104 (156) & 9,339 (37,802,124) & - Not run yet -\\  
[1ex]    % [1ex] adds vertical space 
\hline   % inserts single-line 
\end{tabular} 

{\footnotesize $^1$ Simple, connected, 3-regular, and 3-colourable.\qquad\qquad\qquad\qquad\qquad\qquad\qquad\qquad\qquad\qquad\qquad\qquad\qquad\qquad\,\,\,\,\,}\ 

{\footnotesize $^2$ Hypohamiltonian. Confirmed existence of non-empty $open(V)$ after removing a vertex and re-running the WCA.}\

}\
\end{table}

\footnotetext[8]{Historical Note: Ignoring the planarity condition on Tait's conjecture, the Matteo graph \cite{Leder16Other} is the smallest non-planar counterexample, while the Barnette-Bosák-Lederberg graph \cite{Leder16} is the smallest planar, 3-colourable, 3-connected counterexample to Tait's conjecture (Tutte's graph is a larger counterexample). We also note that the Georges graph is the smallest counterexample to Tutte's conjecture, and Horton's graph was the first counterexample to Tutte's conjecture.}\

\newpage

\section{Discussion}\label{section6} 

\subsection{About Practical Generalizations of the Algorithm}\label{section61} The algorithm can be designed to invoke arbitrary levels of overlay i.e. adaptive strategies that change the level of overlay if more depth is desired / needed to deduce variables at zero level. But in order to make use of increased overlay, it's necessary to add more variables to retain information about tests for matching. For example, if we create a quadrupal overlay version of the algorithm, we then introduce $\{0,1\}$ $\{p_{u,i},p_{v,j},p_{w,k}\}$ variables and redefine system 1 and matrix $Q$ in terms of triply nested Birkhoff polyhedra. See the discussion in \cite{gismo08} for a description of these polyhedra as feasible regions of LP formulations (relaxed IPs). There exists a sequence of feasible regions in correspondence with increasing levels of nested Birkhoff polyhedra whose end feasible region is the convex hull of the set of integer extrema of system 1. See \cite{dg08} for a discussion of facet-inducing inequalities.\

The term closure has so far been reserved for deducing variables added to $E$ by invoking the algorithm. But other polynomial time techniques can be used to deduce variables at zero level. For example, prior to matching, we could implement LP and maximize each variable in system 1, and if its maximum is less than unit level, the variable can be set zero. In our implementation, we use Boolean closure. See \cite{sgsb16} for more details. We also note there exist entire conferences devoted to \textit{Matching Under Preferences} \cite{micro2017}. Perhaps many more innovative heuristics exist and can be included in the algorithm.\

The algorithm is designed for parallel processing. Each $q_{u,i,v,j}$ variable not yet deduced zero can be tested independent of the others by making a copy of matrix $Q$ and implementing the algorithm. If an independent process deduces a $q_{u,i,v,j}$ variable at zero level, simply update the corresponding $q_{u,i,v,j}$ variable in each $Q$ across all processes.\

For some applications, there exist model specific dependencies between variables i.e. undirected HCP implies $\{p_{u,i} ,p_{v,j}\}=0$ if and only if $\{p_{u,n+1-i} ,p_{v,n+1-j}\}=0$. In this way we account for companion cycles.\

\subsection{About Study of the Algorithm}\label{section611} Exclusion set $E$ is the focus of study. We propose to classify different $E$ by the pattern that remains in matrix $Q$ after exit from the algorithm (up to isomorphism) i.e. $Q$ covers the set of all possible solutions to the IP. It would be useful to know what kinds of $E$ cause the algorithm to generate $Q$ as a minimal cover, since it then follows that the algorithm would \textit{decide} feasibility of the IP. Even if there exist classes of $E$ for which infeasible IPs provably exit the algorithm infeasible, no matter that $Q$ is or is not a minimal cover, it still follows that the algorithm \textit{decides} feasibility of the IP.\

We plan to investigate counterexamples via the matching model for HCP. Graph C7-21 (not 3-regular) fails an earlier version of the algorithm \cite{sgsb16}. We will convert \cite{hay2} C7-21 and study it as instance of 3HCP.\

\subsection{Two More Matching Model Applications for Input to the Algorithm}\label{section62} We now present two more matching models as applications for the algorithm\footnotemark[9]. $Q$'s components no longer have the interpretation as sequenced arcs in a cycle. Instead, let $Q$ be an $m^2$x$m^2$ block permutation matrix, whose blocks are $m$x$m$ permutation matrices $P$. We note from \cite{gismo08} that $F$ is a subgraph of $G$ if and only if there exists permutation matrix $P$ such that $P^TGP$ covers $F$ (and we add) if and only if $Q \cdot \textbf{g}$ covers $\textbf{f}$, where $\textbf{f}$ and $\textbf{g}$ are column vectors of adjacency matrices $F$ and $G$ formatted as $\{F(1,1),F(1,2),...,F(1,m),F(2,1),F(2,2),...,F(m,m)\}$ and $\{G(1,1),G(1,2),..., G(1,m),G(2,1),G(2,2),...,G(m,m)\}$.\footnotetext[9]{See \cite{gismo13a} for more information about these modelling techniques.}\

We now model both the graph and subgraph isomorphism decision problems as matching models, the single difference being that in the case of graph isomorphism, more information appears to be added to $E$. First note that $Q \cdot \textbf{g}$ \textit{covers} $\textbf{f}$ means $Q \cdot \textbf{g}$ is required to place ones in the same positions as those of $\textbf{f}$. So for each of these equations, a subset of row components sum to one implying that the complement row components must therefore all be set at zero level. Add them to $E$. This completes the subgraph isomorphism matching model and only part of the graph isomorphism model. For graph isomorphism, \textit{cover} means \textit{equality}. The remaining equations to be satisfied are those for which $Q \cdot \textbf{g}$ is required to place zeroes in the same positions as those of $\textbf{f}$. So for each of these equations, a subset of row components sum to zero implying that these row components must therefore all be set at zero level. Add them to $E$. This completes the graph isomorphism matching model.\

\subsection{Other Applications of the Algorithm}\label{section63} We originally intended for the algorithm to decide feasibility of a matching model. When it decides infeasibility, the algorithm has served its purpose. Otherwise it's not known if the model is feasible or infeasible. We note that $open(V)$ is a refined cover of possible solutions to the IP and we believe that this is useful. We propose that the algorithm can be developed as part of other search based algorithms, either to provide refined information prior to a search, or incorporated and updated alongside a search based algorithm to provide more information during a search.\

There is one last thought about an academic use for the algorithm. Suppose we are given a correctly guessed infeasible IP, and the algorithm exits undecided. We can attribute the failure to $E$ as lacking the necessary / right kind of $\{p_{u,i},p_{v,j}\}$ that could induce closure. We could then theoretically augment $E$ with additional $\{p_{u,i},p_{v,j}\}$ until we deduce infeasibility, and discover extra information needed to generate $open(V)=\emptyset$. So for application, when the algorithm \textit{gets stuck} and $open(V) \ne \emptyset$, simply augment $E$ with additional $\{p_{u,i},p_{v,j}\} \in open(V)$, and test if $open(V)$ becomes empty. While it might be difficult to guess minimal sized sets of additional $\{p_{u,i},p_{v,j}\}$, if they can be guessed, we will then have articulated what critical information is needed to solve the problem. Of course it's not known if these additional $\{p_{u,i},p_{v,j}\}$ can be efficiently computed or validated as members in $E$. See conjecture \ref{con1}.\

\section{Acknowledgements and Dedication}\label{section7} Thank you to: Adrian Lee for preparing and running some of the examples presented in Tables \ref{tab1} and \ref{tab2}; Nicholas Swart for testing and implementing 2000 31-vertex non-Hamiltonian graphs in 2013; Catherine Bell for suggestions and contributions early on in this project.\

We dedicate this paper to the late Pal Fischer (16/11/2016). For Ted, Pal was a colleague, co-author and friend. For myself (Gismondi), Pal taught me analysis, an understanding of convex polyhedra, and later became a colleague. Ted and I both already miss him very much.\

%%%%%%%%%%%%%%%%%%%%%%%%%%%%%%%%%%%%%%%%%%%

%%%%%%%%%%%%%%%%%%%%%%%%%%%%%%%%%%%%%%%%%%%
%\bibliography{ReferenceDB.bib}
%\bibliographystyle{apacite}

\end{document}